\documentclass[doublespacing]{cpcauth}

\usepackage{graphicx}
\usepackage{amsmath,amssymb}

\begin{document}

\begin{frontmatter}

\title{Modelling nematohydrodynamics in liquid crystal devices}

\author{G\'eza T\'oth$^a$, Colin Denniston$^b$ and J.M. Yeomans$^a$}

\address{$^{a}$ Dept. of Physics, 
University of Oxford, Theoretical Physics, 1 Keble Road, Oxford OX1
3NP, UK}
\address{$^{b}$ Department of Physics and Astronomy, The Johns Hopkins
  University, Baltimore, Maryland 21218, USA}

\begin{abstract}

\corauth[1]{E-mail:yeomans@thphys.ox.ac.uk, 
phone: +44 (0)1865 273992,
fax:   +44 (0)1865 273947.}

We formulate a lattice Boltzmann algorithm which solves the hydrodynamic
equations of motion for nematic liquid crystals. The applicability of
the approach is demonstrated by presenting results for two liquid
crystal devices where flow has an important role to play in the switching.
\end{abstract}
\end{frontmatter}

\section{Introduction}

In this paper we describe a lattice Boltzmann algorithm which solves
the hydrodynamic equations of motion for nematic liquid
crystals. Liquid crystals are widely used in display devices and, as
examples of applications of the algorithm, 
we consider two model devices where flow
has an important role to play in the switching.

Liquid crystals are fluids made up of rod-shaped molecules\cite{dG93}. At high
temperatures and low concentrations the angular distribution of the
molecules is isotropic. However, as the temperature is lowered or the
concentration increased, a liquid crystal can undergo a phase
transition to a state where the molecules tend to align parallel, the
so-called nematic phase. Liquid crystals exhibit complicated,
non-Newtonian flow behaviour because of the coupling between this
molecular structure and the flow field\cite{L88}. 
Examples include shear
thinning and non-equilibrium phase transitions such as banding under
shear\cite{OL99}.

Another consequence of the ordering of nematic liquid
crystals is their widespread use in optical display devices such
as those on digital watches or calculators. 
In a typical display device, the liquid crystal is confined between
two plates a few microns apart. The molecular configuration on the
plates is fixed.  When an electric field is switched on molecules in
the bulk
align in the direction preferred by the field.  After switching off
the field, long-range elastic interactions ensure that the molecules
reorient themselves in the direction preferred by the surfaces. 
These devices can be used as displays because different liquid crystal
orientations have different optical properties. Flow can be important
in device operation because it can control the speed of switching or
even select the states between which switching can occur.

Modelling flow in non-Newtonian fluids such as liquid crystals 
is difficult because of the
need to incorporate the coupling between the microscopic structure and
the flow. The hydrodynamic equations of motion can be extremely
complex and numerical solutions demanding.  A method which has
recently proved very successful in modelling complex fluids is
the lattice Boltzmann approach\cite{CD98}. Lattice
Boltzmann simulations solve the hydrodynamic equations
of motion while inputting sufficient, albeit generic, molecular information to
model the important physics of a given fluid. This is often
done by imposing a Landau free energy functional, so the fluid
evolves to a known thermodynamic equilibrium\cite{S96}. The method
has been successfully used to investigate domain growth in binary
mixtures and liquid-gas systems\cite{Y99}, the flow of binary mixtures
in porous media, and ordering in amphiphilic fluids\cite{LG99}.

Here we construct a lattice Boltzmann algorithm to solve the
hydrodynamic equations of motion for nematic liquid crystals.
We follow Beris and Edwards\cite{BE94} and 
consider a rather general formalism of the
hydrodynamics written in terms of a tensor order 
parameter, $\bf Q$.
This approach lends itself to a lattice Boltzmann interpretation: the
partial density distribution functions which are the usual variables
in the simulations are supplemented by a set of tensors 
related to $\bf Q$. Backflow,
the hydrodynamics of topological defects and the possibility of 
transitions between
the nematic and isotropic phases appear naturally within the formalism.

The next section of the paper summarises the Beris-Edwards equations of motion
for liquid crystal hydrodynamics.
Sec. 3 describes the lattice Boltzmann algorithm.  General viscosity
terms, needed to match to experimental viscosities, are included.
The approach is applied to switching in
a Frederiks cell in Sec. 4 and, in Sec. 5, used to confirm that backflow
can influence state selection in a bistable liquid crystal device.

\section{Hydrodynamic equations of motion}

Ordering in a nematic liquid crystal can be described in terms of a
tensor order parameter ${\bf Q}$ which is symmetric and traceless. 
Expanding the free energy in terms of the order parameter gives 
the Landau-de-Gennes expression\cite{dG93}
\begin{equation}
{\mathfrak F}=\int_V dV \left\{F_b-F_E+F_d \right\}+\int_{\partial V} dS
\left\{F_a\right\},
\end{equation}
where
\begin{eqnarray}
F_b&=&\frac{A}{2} (1-\frac{\gamma}{3})
  Q_{\alpha \beta}^2 - \frac{A \gamma}{3} Q_{\alpha \beta}
Q_{\beta \gamma}Q_{\gamma \alpha} \nonumber\\
&& +\frac{A \gamma}{4}(Q_{\alpha \beta}^2)^2,\nonumber\\
F_{d} &=&\frac{L_1}{2} (\partial_\alpha 
               Q_{\beta \gamma})^2+
               \frac{L_2}{2} (\partial_\alpha 
               Q_{\alpha \gamma})
               (\partial_\beta 
               Q_{\beta \gamma}) \nonumber \\
&+&\frac{L_3}{2} Q_{\alpha \beta}
              (\partial_\alpha Q_{\gamma \epsilon})
               (\partial_\beta Q_{\gamma \epsilon}),\nonumber\\
F_a&=&\frac{\alpha_s}{2} (Q_{\alpha \beta}-Q^{0}_{\alpha \beta})^2, \;\;
F_E=\frac{\epsilon_a}{12 \pi}E_{\alpha}Q_{\alpha\beta}E_{\beta}.
\label{free}
\end{eqnarray}
(Greek subscripts represent Cartesian directions and the usual
summation over repeated indices is assumed.)  The bulk free energy
terms $F_b$ describe a liquid crystal with a
first-order, isotropic--nematic transition at $\gamma=2.7$\cite{DE89}.
Contributions to the free energy which arise from an imposed
electric field $\vec{E}$ are included in $F_E$.  $F_d$ describes the
elastic free energy\cite{BE94} and
$F_a$ is a surface free energy which fixes a
preferred orientation ${\bf Q}^{0}$ for the surface director field.

${\bf Q}$  evolves according to a convection-diffusion equation\cite{BE94}
\begin{equation}
(\partial_t+{\vec u}\cdot{\bf \nabla}){\bf Q}-{\bf S}({\bf W},{\bf
  Q})= {\Gamma} {\bf H}
\label{Qevolution}
\end{equation}
where ${\vec u}$ is the bulk fluid velocity and ${\Gamma}$ 
is a collective rotational diffusion constant.
The term on the right-hand side of Eqn.\ (\ref{Qevolution})
describes the relaxation of the order parameter towards the minimum of
the free energy ${\mathfrak F}$,
 \begin{equation}
{\bf H}= -{\delta {\mathfrak F} \over \delta {\bf Q}}+({\bf
    I}/3) {\mbox Tr}\left\{ \delta {\mathfrak F} \over \delta {\bf Q} \right\}.
\label{H(Q)}
\end{equation}  
The order parameter distribution can be both rotated and stretched by
flow gradients.  This is described by the term on the left-hand side
\begin{eqnarray}
&&{\bf S}({\bf W},{\bf Q})
=(\xi {\bf D}+{\bf \Omega})({\bf Q}+{\bf I}/3)+\\
&&\qquad({\bf Q}+{\bf I}/3)(\xi {\bf D}-{\bf \Omega})
-2 \xi ({\bf Q}+{\bf I}/3){\mbox{Tr}}({\bf Q}{\bf W})\nonumber
\end{eqnarray}
where  ${\bf D}=({\bf W}+{\bf W}^T)/2$ and
${\bf \Omega}=({\bf W}-{\bf W}^T)/2$
are the symmetric part and the anti-symmetric part respectively of the
velocity gradient tensor $W_{\alpha\beta}=\partial_\beta u_\alpha$ and
$\xi$ is related to the aspect ratio of the molecules. 

The flow of the liquid crystal fluid obeys the continuity and
Navier-Stokes equations
\begin{equation}
\partial_t \rho + \partial_{\alpha} \rho u_{\alpha}= 0,
\label{cont}
\end{equation}
\begin{eqnarray}
& &  \rho\partial_t u_\alpha+\rho u_\beta \partial_\beta
u_\alpha=\partial_\beta \sigma_{\alpha\beta}+
\label{ns}\\
& & \quad\,\frac{\rho \tau_f}{3}\partial_\beta((\delta_{\alpha\beta}+3\partial_n
\sigma_{\alpha\beta})\partial_\zeta u_\zeta+\partial_\alpha
u_\beta+\partial_\beta u_\alpha)
\nonumber
\end{eqnarray}
where 
\begin{eqnarray}
\sigma_{\alpha\beta} &=&-\rho T \delta_{\alpha \beta}
-(\xi - 1)H_{\alpha\gamma}(Q_{\gamma\beta}+{1\over
  3}\delta_{\gamma\beta})\nonumber\\
&-&(\xi+1) (Q_{\alpha\gamma}+{1\over
  3}\delta_{\alpha\gamma})H_{\gamma\beta}\nonumber\\
&+& 2\xi (Q_{\alpha\beta}+{1\over 3}
\delta_{\alpha\beta})Q_{\gamma\epsilon}
H_{\gamma\epsilon}
- \partial_\beta Q_{\gamma\nu} {\delta
{\mathfrak F}\over \delta\partial_\alpha Q_{\gamma\nu}}\nonumber\\
&+& \beta (Q_{\alpha\beta}+{1\over
  3}\delta_{\alpha\beta})Q_{\gamma\epsilon}W_{\gamma\epsilon}+\sigma_{M,\alpha\beta}
\label{stress}
\end{eqnarray}
is the stress due to the nematic order\cite{BE94} and
the external electric field.
The latter is represented by the $\sigma_{M,\alpha\beta}$
Maxwell stress tensor\cite{LL60}.

\section{Lattice Boltzmann algorithm}

A lattice Boltzmann scheme which reproduces Eqns.~(\ref{Qevolution}), 
(\ref{cont}) and (\ref{ns}) to 
second order can be defined in terms of two distribution
functions $f_i (\vec{x})$ and ${\bf G}_i (\vec{x})$ where $i$ labels
lattice directions from site $\vec{x}$. Physical variables are related to 
the distribution functions by
\begin{equation}
\rho=\sum_i f_i, \qquad \rho u_\alpha = \sum_i f_i  e_{i\alpha},
\qquad {\bf Q} = \sum_i {\bf G}_i.
\end{equation}

The distribution functions evolve in a time step $\Delta t$ according to
\begin{eqnarray}
&&(f_i({\vec x}+{\vec e}_i \Delta t,t+\Delta t)-f_i({\vec x},t))/\Delta t=
\label{evol}\\
&&\,\frac{1}{2} \left[{\mathcal C}_{fi}({\vec x},t,\left\{f_i \right\})+ {\mathcal C}_{fi}({\vec x}+{\vec e}_i \Delta t,t+\Delta
t,\left\{f_i^*\right\})\right],
\nonumber\\ 
& &({\bf G}_i({\vec x}+{\vec e}_i \Delta t,t+\Delta t)-{\bf G}_i({\vec
  x},t))/\Delta t=\nonumber\\
&&\,\,\frac{{\mathcal C}_{{\bf G}i}({\vec
    x},t,\left\{{\bf G}_i \right\})+{\mathcal C}_{{\bf G}i}({\vec
    x}+{\vec e}_i \Delta t,t+\Delta t,\left\{{\bf G}_i^*\right\})}{2}
\nonumber
\end{eqnarray}
where the collision operators are taken to have the form of a single
relaxation time Boltzmann equation, together with a forcing term
\begin{eqnarray}
& &{\mathcal C}_{fi}({\vec x},t,\left\{f_i \right\})= 
\label{collision}\\
& &\quad-\frac{f_i({\vec x},t)-f_i^0({\vec x},t,\left\{f_i \right\})}{\tau_f}
+p_i({\vec x},t,\left\{f_i \right\}),\nonumber\\ 
& &{\mathcal C}_{{\bf G}i}({\vec x},t,\left\{{\bf G}_i \right\})=\nonumber\\ 
& &\quad-\frac{{\bf G}_i({\vec x},t)-{\bf G}_i^0({\vec x},t,\left\{{\bf G}_i \right\})}{\tau_{\bf G}}
+h_i({\vec x},t,\left\{{\bf G}_i \right\}).
\nonumber
\end{eqnarray}
$f_i^*$ and ${\bf G}_i^*$ in equations (\ref{evol}) and (\ref{collision})
are first order approximations to $f_i({\vec x}+{\vec e}_i \Delta
t,t+\Delta t)$ and ${\bf G}_i({\vec x}+{\vec e}_i \Delta t,t+\Delta
t)$. They are introduced to remove lattice viscosity terms to second
order and they give improved stability.

The form of the equations of motion and thermodynamic equilibrium
follow from the choice of the moments of the equilibrium distributions
$f^0_i$ and ${\bf G}^0_i$ and the driving terms $p_i$ and
$h_i$. $f_i^0$ is constrained by
\begin{eqnarray}
&&\sum_i f_i^{0} = \rho,\qquad \sum_i f_i^{0} e_{i \alpha} = \rho
u_{\alpha}, \nonumber\\ 
&&\sum_i f_i^{0} e_{i\alpha}e_{i\beta} = -\sigma_{\alpha\beta}^s+\rho
u_\alpha u_\beta
\label{mom1} 
\end{eqnarray}
where the zeroth and first moments are chosen to impose conservation of
mass and momentum. The second moment of $f^{0}$ controls the symmetric
part of the stress tensor, whereas the moments of $p_i$
\begin{eqnarray}
&&\sum_i p_i = 0,  \sum_i p_i e_{i\alpha} = \partial_\beta
\sigma_{\alpha\beta}^a, \nonumber\\
&&\sum_i p_i
e_{i\alpha}e_{i\beta} = 0
\end{eqnarray}
impose the antisymmetric part of the stress tensor.

For the equilibrium of the order parameter distribution we choose
\begin{eqnarray}
 &&\sum_i {\bf G}_i^{0} = {\bf Q},\quad \sum_i
{\bf G}_i^{0} {e_{i\alpha}} = {\bf Q}{u_{\alpha}},\nonumber\\
&&\quad \sum_i {\bf G}_i^{0}
e_{i\alpha}e_{i\beta} = {\bf Q} u_\alpha u_\beta .
\end{eqnarray}
This ensures that the fluid minimizes its free energy at equilibrium
and that it is convected with the flow. Finally the evolution of the
order parameter is most conveniently modelled by choosing
\begin{eqnarray}
 &&\sum_i h_i =\Gamma {\bf H}({\bf Q})
+{\bf S}({\bf W},{\bf Q}),\nonumber\\ 
&&\qquad \sum_i h_i {e_{i\alpha}} = (\sum_i h_i) {u_{\alpha}}.
\label{mom4}
\end{eqnarray}
Conditions (\ref{mom1})--(\ref{mom4}) can be satisfied by taking
$f_i^{0}$, ${\bf G}_i^{0}$, $h_i$, and $p_i$ as polynomial
expansions in the velocity as is usual in lattice Boltzmann
schemes\cite{CD98}.
A second order Chapman--Enskog expansion for the evolution equations 
(\ref{evol}) incorporating the conditions
(\ref{mom1})--(\ref{mom4}) leads to the equations of 
motion\cite{DO01}.

\section {Frederiks cell}

\begin{figure}
\begin{center}
\includegraphics*[width=6cm]{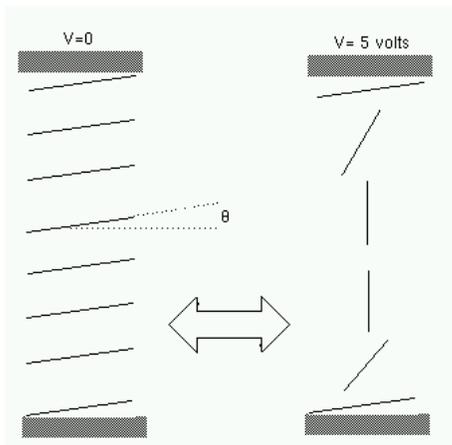}
\end{center}
\caption{The two stable states of a Frederiks cell.  At zero voltage
  the liquid crystal prefers the orientation imposed by the boundaries.  When
  the electric field is turned on the nematic prefers to align 
 along the field direction.}
\label{fred}
\end{figure}

One of the simplest liquid crystal devices is the Frederiks cell depicted in
Fig.~\ref{fred}. 
When the voltage is increased above a critical value $V_c$ switching
from the zero voltage state occurs rather rapidly.  The factor limiting
the speed at which the device can be updated is the switching from the
voltage aligned state to that preferred by the boundaries.  This
switching involves both relaxation of the free energy and viscous
dissipation associated with hydrodynamic modes\cite{dG93}. Our aim is to
quantitatively model this phenomenon for a given set of
elastic constants and viscosities\cite{B72}. 

\begin{figure}
\begin{center}
\includegraphics*[width=6cm]{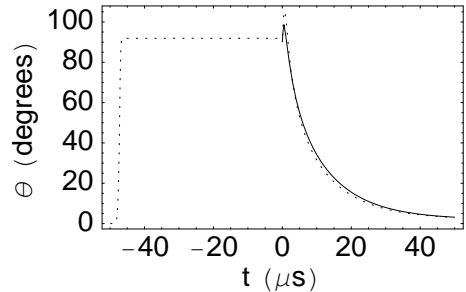}
\end{center}
\caption{Angle $\theta$, with the horizontal, of the
director at the centre of the cell as a
  function of time.  The system starts in the fully relaxed,
zero voltage state and the voltage is turned on immediately at the far
left of the plot.  The voltage is turned off at $t=0$.}
\label{fredtheta}
\end{figure}

Figure~\ref{fredtheta} shows $\theta$, the director
angle\cite{director} measured in the
centre of the channel, as indicated in Fig.~\ref{fred}, as a function of
time as the voltage is switched on and then off.  The parameters used
for the simulation are listed in footnote~\cite{foot1} and were chosen
to correspond to commercially relevant materials.
For comparison, we show relaxation data, provided by Sharp
Laboratories of Europe, taken from a simulation using a
commercial code which solves the Ericksen-Leslie\cite{EL} equations
in one dimension. (For a similar,
but not identical, Ericksen-Leslie calculation and comparison to 
experimental data see
Ref.~\cite{KJ99}.)  For the most part, there is very good agreement between
the two methods.

The ``blip'' seen in the plot when the voltage is turned off is a
real effect known as the optical bounce, due to its optical signature.
After the removal of the electric field, the changing director field
induces a hydrodynamic flow.  This, in turn, couples back to the
director field and causes it to
momentarily move in the ``wrong'' direction.  Viscous dissipation in
the fluid rapidly damps out the flow and the director can then relax
toward the minimum of the free energy.    The bounce we see is somewhat
higher than that observed in the Ericksen-Leslie simulations.
However recent attempts to exactly
match the quantitative features of the optical bounce to solutions of the
Ericksen-Leslie equations have suggested that they always
underestimate its value relative to experiments\cite{KJ99}.  Thus,
the additional details in the tensor model may be better
reproducing the experimental results.

\section {Flow induced surface switching}

A current aim of the device industry is to construct bistable displays
which can remain in two different (meta)stable states in zero
field. This would, for example, lead to significant power saving in
infrequently updated displays or could be used in the construction of
smart cards which rely on an external power source.

\begin{figure}
\begin{center}
\includegraphics*[width=1cm]{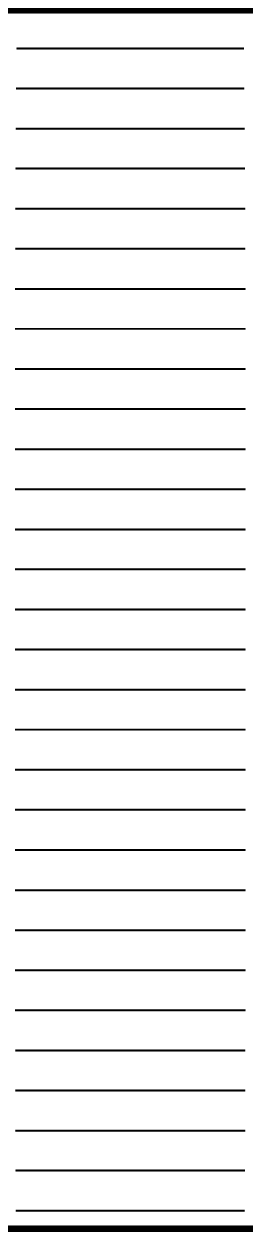}
\hskip 1true cm
\includegraphics*[width=1cm]{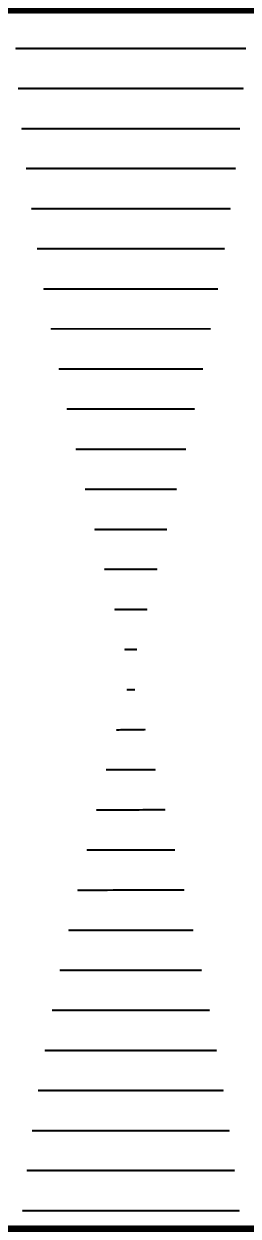}
\vskip 0.2true cm
(a) {\hskip 1.5cm} (b)
\end{center}
\caption{The two stable states
(a) horizontal and (b) twist 
of the surface switching device considered in Sec.~5. In (b) the molecular
orientation twists to perpendicular to the plane of the figure at the
middle of the sample.}
\label{fig_twostates}
\end{figure}

\begin{figure}
\begin{center}
\includegraphics*[width=1.5cm]{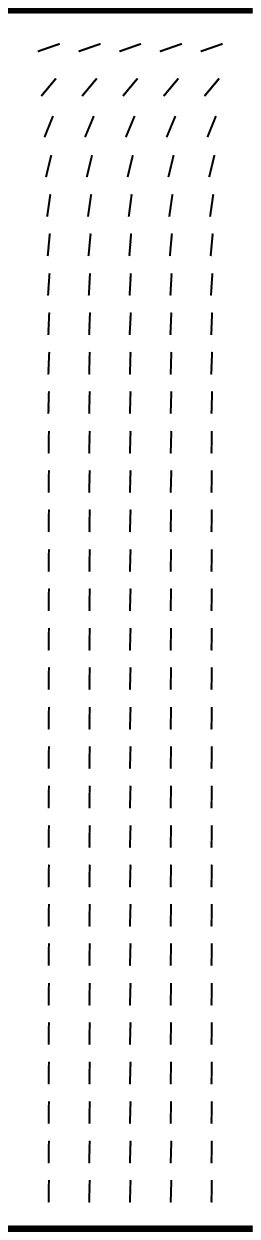}
\hskip 1true cm
\includegraphics*[width=1.5cm]{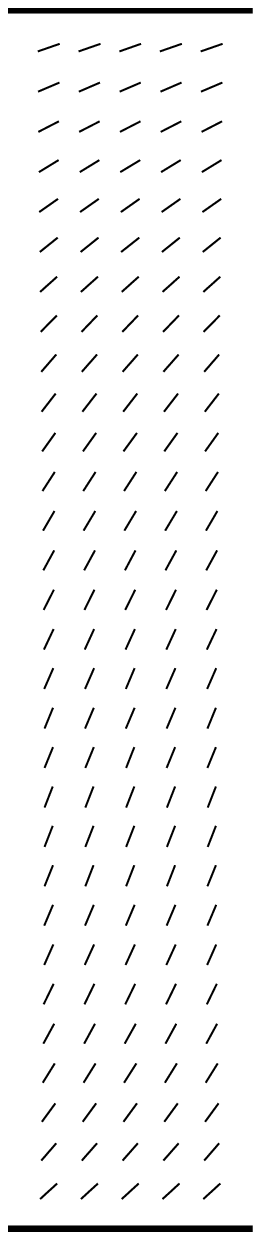}
\hskip 1true cm
\includegraphics*[width=1.5cm]{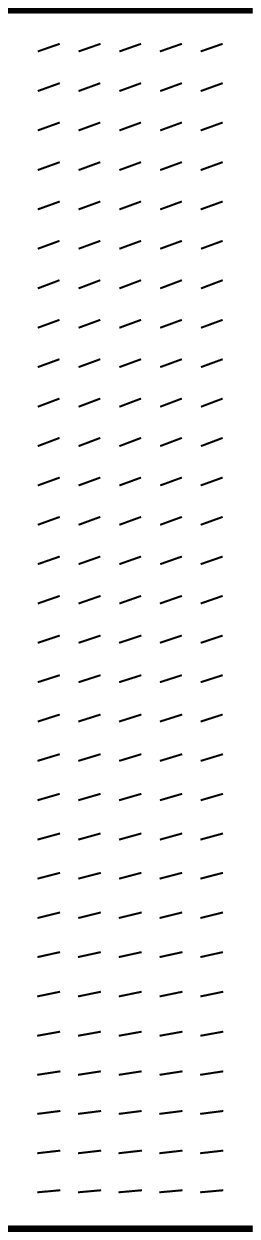}
\vskip 0.2true cm
(a) {\hskip 1.9cm} (b) {\hskip 1.9cm} (c)
\end{center}
\caption{Snapshots of the director configuration of the surface
switching device after the electric field
is switched off, but back-flow effects are not incorporated in the model.
The (a) vertical state relaxes through (b) an intermediate state 
to (c) the horizontal state. The snapshots correspond to $0$, $7$ and 
$15$ ms.
At the top surface 
the molecules are strongly pinned at a
$20^o$ pretilt to the horizontal axis. Pinning at the
bottom surface is weak with a $0^o$ pretilt.} 
\label{fig_switch_nf}
\end{figure}

In this paper we will consider one possible mechanism for accessing
two zero-field states, flow induced surface switching\cite{DN97}.
The two states are the horizontal and the twist states
shown in Fig. \ref{fig_twostates}. 
The liquid crystal is confined between two plates. The top
(bottom) plate has strong (weak) anchoring.
The horizontal state can be switched to the vertical state by
an electric field. Then, if the field is abruptly switched off,
one might expect the system to relax back to the horizontal state. 
The relaxation
commences at the boundaries, where the molecules start to rotate in
the same direction at the top and bottom surfaces. If one does not consider
the effect of the back-flow (equivalent to using a Ginzburg-Landau 
equation to describe the dynamics), then this is indeed the case, as
shown by the simulation results in Fig. \ref{fig_switch_nf}.
(The simulation parameters are summarized in \cite{SYMPARAM}.
Back-flow is switched off by setting the velocity to 
zero in Eqn.~(\ref{Qevolution}).

\begin{figure}
\begin{center}
\includegraphics*[width=1.5cm]{cpc2001_bistable_vertical.eps}
\hskip 1true cm
\includegraphics*[width=1.5cm]{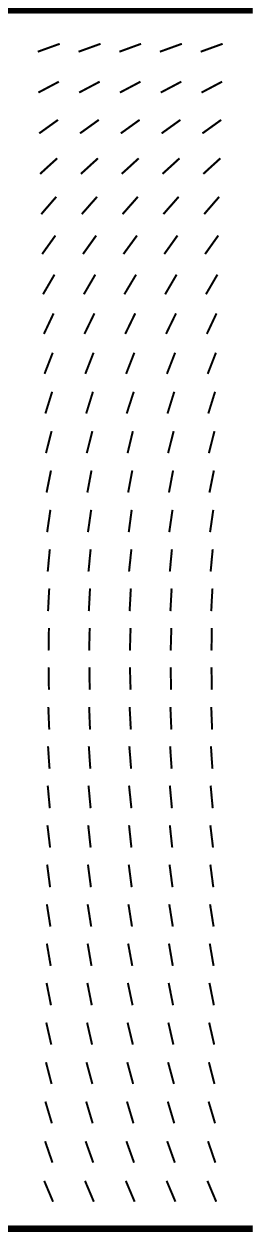}
\hskip 1true cm
\includegraphics*[width=1.5cm]{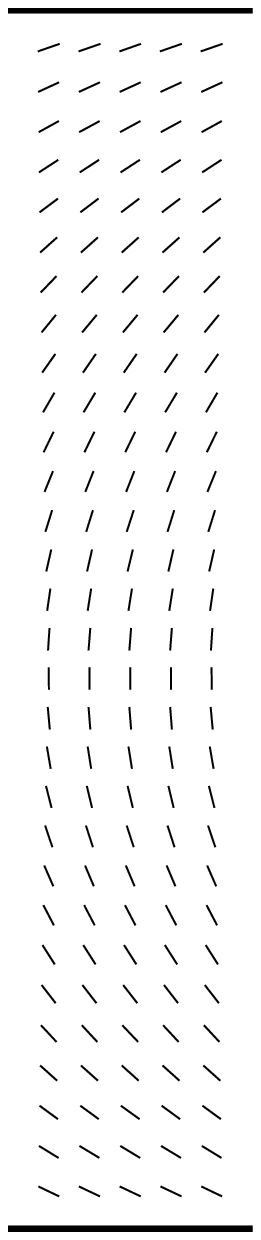}
\vskip 0.2true cm
(a) {\hskip 2cm} (b){\hskip 2cm} (c)
\end{center}
\caption{
Snapshots of the director configuration of the surface switching
device after the electric field
is switched off, and back-flow effects are incorporated in the model.
The (a) vertical state 
relaxes through (b) an intermediate state 
to (c) the bend state. The snapshots correspond to $0$, $1$ and 
$6$ ms.}
\label{fig_switch}
\end{figure}

If, however, back-flow is included, a horizontal flow is created
starting from the top plate and gradually reaching the 
bottom surface. This horizontal flow causes the molecules 
to rotate in the opposite direction at the bottom surface.
As a result the vertical state transforms to
a bend state, as shown in
Fig. \ref {fig_switch}. 
There is an energy argument which suggests that the bend 
state is unstable and quickly relaxes to a twist state\cite{DN97}
if
the Frank elastic constant $K_{22}$ is 
sufficiently small compared to $K_{11}$ and $K_{33}$.

Hence in a bistable device, if the field is switched off gradually, 
the effect of back-flow is weak, and the liquid crystal settles in the
horizontal
state. If the field is switched off abruptly, then it settles in the twist
state. When a field is reapplied both zero field configurations
switch back to the vertical state.

\section {Conclusion}

In conclusion, we have derived and implemented a lattice Boltzmann
algorithm for liquid crystal hydrodynamics.  This opens the way to
investigate a wide range of physical phenomena which result from the
coupling between the director field and the flow, many of which have
been examined only indirectly or with severe approximations
in the past. Two examples have been presented showing the relevance of
flow to switching in liquid crystal devices.\\
~\\
We thank E. Acosta for helpful conversations. Financial support from
the EPSRC is acknowledged.


\begin{thebibliography}{00}

\bibitem{dG93}
P.G. de Gennes and J. Prost, The Physics of Liquid Crystals, 2nd
  Ed. (Clarendon Press, Oxford , 1993). 

\bibitem{L88}
R.G. Larson, Constitutive Equations for polymer melts and
  solutions (Butterworths series in Chem. E., Butterworth, 1988).

\bibitem{OL99}
P.D. Olmsted and C-Y. David Lu, Phys. Rev. E 60 (1999) 4397.

\bibitem{CD98}
S. Chen and G. D. Doolen, Annual Rev. Fluid Mech. 30 (1998) 329.

\bibitem{S96}
M.R. Swift, E. Orlandini, W.R. Osborn and J.M. Yeomans, Phys. Rev. E
54 (1996) 5041.

\bibitem{Y99}
J.M. Yeomans, Ann. Rev. Comp. Phys. VII (1999) 61.

\bibitem{LG99}
A. Lamura, G. Gonnella and J.M. Yeomans, Europhys. Lett. 45 (1999) 314; 
O. Theissen and G. Gompper, European Phys. J. B11 (1999) 91.

\bibitem{BE94}
A.N. Beris and B.J. Edwards, Thermodynamics of Flowing Systems
(Oxford University Press, Oxford, 1994).

\bibitem{DE89}
M. Doi and S. Edwards, The Theory of Polymer Dynamics (Clarendon
Press, Oxford, 1989).

\bibitem{LL60}
L.D. Landau and E.M. Lifshitz, Electrodynamics of continuous
(Pergamon Press, New York, 1960).

\bibitem{DO01} C. Denniston, E. Orlandini and J. M. Yeomans,
Phys. Rev. E 63 (2001) 056702.

\bibitem{B72}
F. Brochard, E. Guyon, and P. Pieranski, Phys. Rev. Lett. 28 (1972) 1681.

\bibitem{director}
The director field is an order parameter giving the average local
direction of the molecules. It lies along the largest eigenvector of
${\bf Q}$.

\bibitem{foot1}
The parameters used correspond to: a starting voltage  5.6214 V, 0 V
during relaxation, $\epsilon_\parallel=19$, $\epsilon_\perp=5.2$,
$\gamma_1=0.15$ Pa s; Meisowicz viscosities $\eta_1=0.17$ Pa s,
$\eta_2=0.04$ Pa s, $\eta_3=0.04$ Pa s, $\eta_{12}=0$; elastic constants
$K_{11}=11.1$ pN, $K_{22}=6.7$ pN, $K_{33}=17.1$ pN; pretilt at
surface $2 ^\circ$, cell thickness $3 \mu$m.

\bibitem{EL}
The Ericksen-Leslie equations (J.L. Ericksen, Phys. Fluids 9
(1966) 1205; F.M. Leslie, Arch. Ration. Mech. Analysis 28 
(1968) 265.) are hydrodynamic equations of motion for
nematics written in terms of the director field. The Beris-Edwards
equations considered in this paper reduce to them for a uniaxial order
parameter of constant magnitude. 

\bibitem{KJ99}
J. Kelly, S. Jamal, and M. Cui, J. Appl. Phys. 86 (1999) 4091.

\bibitem{DN97}
 I. Dozov, M. Nobili and G. Durand, Appl. Phys. Lett. 70 (1997) 1179;
 J. G. McIntosh and F. M. Leslie, J. Eng. Math. 37 (2000) 129.

 \bibitem{SYMPARAM}
Parameters used correspond to:
  sample width $1.7\mu$m;
  $\Gamma=3.33$ Pa$^{-1}$s$^{-1}$,
  $L_{1}=$ 8.75 pN, $L_{2}=$ 6.36 pN and $L_{3}=$ 13.60 pN
(or equivalently 
  $K_{11}=$ 4.83 pN, $K_{22}=$ 3.24 pN, $K_{33}=$ 8.23 pN); 
  Miesowicz viscosities
  $0.02$ Pa s to $0.15$ Pa s.



\end{thebibliography}
\end{document}